\documentclass[pre,aps,twocolumn,showpacs,showkeys]{revtex4}

\usepackage{color}
\usepackage{graphicx}
\usepackage{dcolumn}
\usepackage{amsmath}
\usepackage{amsfonts}
\usepackage{amssymb}
\usepackage{hyperref}

\newcommand{\op}[1]{#1}
\newcommand{\tmtextbf}[1]{{\bfseries{#1}}}
\newcommand{\tmtextit}[1]{{\itshape{#1}}}

\newcommand{\ket}[1]{\text{\ensuremath{\lvert #1 \rangle}}}
\newcommand{\bra}[1]{\text{\ensuremath{\langle #1 \rvert}}}
\newcommand{\av}[1]{{\langle} #1 {\rangle}}

\newcommand{\ketbra}[1]{{\lvert} #1 {\rangle}{\langle} #1{\rvert}}
\renewcommand{\op}[1]{\hat{#1}}

\begin{document}

\title{A competitive game whose maximal Nash-equilibrium payoff requires quantum resources for its achievement}

\author{Charles D. Hill}
\email{cdhill@unimelb.edu.au}
\author{Adrian P. Flitney}
\email{aflitney@unimelb.edu.au}
\affiliation{School of Physics, University of Melbourne, Parkville, VIC 3010, Australia}
\author{Nicolas C. Menicucci}
\email{nmenicucci@perimeterinstitute.ca}
\affiliation{Perimeter Institute for Theoretical Physics, Waterloo, Ontario N2L 2Y5, Canada}


\begin{abstract}
While it is known that shared quantum
entanglement can offer improved solutions to a number of purely
cooperative tasks for groups of remote agents,
controversy remains regarding the legitimacy of quantum games in a competitive setting---in
particular, whether they offer any advantage beyond what is achievable
using classical resources. 
We construct a competitive game between four players based on the
minority game where the maximal Nash-equilibrium payoff when played
with the appropriate quantum resource is greater than that obtainable
by classical means, assuming a local hidden variable model.
The game is constructed in a manner analogous to a Bell inequality.
This result is important in confirming the legitimacy of quantum games.
\end{abstract}

\maketitle

\section{Introduction}\label{sec:intro}

Game theory is a branch of mathematics dealing with strategic interactions of
competitive agents where the outcome is contingent upon the combined actions of
the agents.
In 1999, game theory was formally extended into the quantum realm
by replacing the classical information with qubits and the player actions by
quantum operators~{\cite{Eisert1999,Meyer1999}}.
Since then much work has been done in the new discipline of quantum game theory~{\cite{Flitney2008,Guo2008}}
and attempts have been made to put it on a more formal footing~\cite{Gutoski2006}.
There have been objections that quantum games
are not truly quantum mechanical and have little to do with the underlying
classical games~{\cite{vanEnk2000,vanEnk2002,Levine2005}}.
However, attempts have been
made to counter these arguments~{\cite{Meyer2000}}.
In addition, quantum games have been
shown to be more efficient than classical games,
in terms of information transfer,
and that finite classical games are a proper subset of quantum games~{\cite{Lee2003}},
thus demonstrating that not all quantum games can be reduced to classical ones.
Quantum game protocols have also been proposed that use the non-local
features of quantum mechanics which have no classical analogue~{\cite{Iqbal2005,Iqbal2006}}.

In the present work we construct a game
that is a minimal quantum generalization of a possible classical game
and that has a Nash equilibrium that is not achievable
by any classical hidden variable model.
Our model is distinguished by its competitive nature from situations,
also referred to in the literature as quantum games,
that involve a number of agents solving a cooperative task
by quantum means~{\cite{Cleve2004,Brassard2005}}.

The minority game was introduced in 1997~{\cite{Challet1997}} as a simple
multi-agent model that is able to reproduce much of the behaviour of financial
markets. The agents independently select one of two choices (`buy' or `sell')
and those in the minority win, the idea being that when everyone is buying
prices are inflated and it is best to be a seller and vice versa. In a
one-shot minority game the best players can do is to select among the
alternatives at random with an unbiased coin. The simplest non-trivial
situation is the four player game: only one player can win; however, there is
a fifty percent chance that there is no minority, in which case all players
receive zero payoff. Versions of the minority game utilizing quantum resources
have attracted attention since the probability of the no-minority case can be
eliminated in the four player game~{\cite{Benjamin2001}}, or reduced for even
$N > 4$~{\cite{Chen2004}}. These result are robust even in the presence of
decoherence~{\cite{Flitney2007}}. In addition, utilizing a particular set of
tunable four-party entangled states as the quantum resource shared by the
players, there is an equivalence between the optimal game payoffs and the
maximal violation of the four-party MABK-type Bell
inequality~{\cite{Mermin1990,Ardehali1992,BK93}} for the initial
state~{\cite{Flitney2009}}.

In the quantum minority game each player receives one qubit from a known
entangled state. They can act on their qubit with a local unitary operator,
the choice of which is their strategy. The qubits are then measured in the
computational basis, and payoffs are awarded as in the classical game.
Starting with the GHZ state $( \ket{0000} + \ket{1111}) / \sqrt{2}$, if each
player operates with
\begin{eqnarray}
  \label{eq:NE4} \op{s} = \frac{1}{\sqrt{2}} \left( \begin{array}{cc}
    e^{i \pi / 8} & ie^{- i \pi / 8}\\
    ie^{i \pi / 8} & e^{- i \pi / 8}
  \end{array} \right),
\end{eqnarray}
the resulting superposition contains only those states where one of the four
player is in the minority and so the average payoff $\av{\$}$ is
$\frac{1}{4}$, compared with $\frac{1}{8}$ for the classical
game~{\cite{Benjamin2001}}. When all players use this strategy the result is
a Nash equilibrium, a strategy profile from which no player can improve their
payoff by a unilateral change in strategy. The result is also Pareto optimal,
one from which no player can improve their payoff without someone else being
worse off.

One complaint, however, that can be leveled at the quantum versions is that
the same outcome can be achieved by a purely classical local hidden variable
model. For example, in a four-player minority game a trusted third party could
choose one of the eight classical messages $0001$, $0010$, $0100$, $1000$,
$1110$, $1101$, $1011$, or $0111$ at random and then inform each of the
players of their selected value. None of the players has an incentive to vary
their choice, and the expected payoff is fair to all players. Such an
arrangement would also yield $\av{\$} = \frac{1}{4}$.

In this paper we introduce a competitive game having a Nash-equilibrium
maximal payoff that requires the use of quantum resources. In other words,
this payoff cannot be achieved by players who only have access to classical
resources (i.e., resources whose statistical properties can be modeled using
local hidden variables). This is in contrast to previous work on cooperative
games such as the XOR game, odd cycle game, and magic square
game~{\cite{Cleve2004}}: even though those games were also shown to be
equivalent to a corresponding Tsirelson-type inequality and can be used to
demonstrate a Bell inequality, the critical distinction is that the game we
consider has a competitive aspect, and it is therefore essential to consider
the Nash equilibrium.

\section{Definition of the game}

We now define the game that is the subject of this paper. This four-player
game will be based partly on the minority game and partly on what we call the
\tmtextit{anti-minority game}. While the minority game provides a payoff of~1
for the player who answers differently from the other three (if there is
exactly one such player) and no payoff to any other player, the anti-minority
game rewards the case where there is no minority, providing a payoff of
$\frac{1}{4}$ to all players when all players give the same answer or there is
a 50/50 split. That is, all the players score $\frac{1}{4}$ on just those
occasions when there would be no winner in a minority game.

The overall game is a combination of these two games.  The players do not know beforehand whether the payoff matrix will be that of the minority game or that of the anti-minority game.  (The way it is selected will be described shortly.)  The players are allowed to meet privately before the game to
agree on a joint strategy and, if they wish, prepare physical resources
(classical or quantum) for each of them to bring with them to the game.  The
players are subsequently isolated and prevented from communicating for the
rest of the game.  An impartial referee (someone other than the players) then asks each of the isolated players one of two questions: either, ``What is the value of $X$?''\ or, ``What is the value of $Z$?''\ to which the player
must respond with either $+1$ or $-1$ as she chooses.  Each player may, if
she wishes, use whatever physical resource she brought with her to aid in
answering her question~\footnote{Cell phones and other communication
devices, which are certainly physical resources, are permitted but are useless
due to the no-communication requirement, which can always be enforced in
principle by sufficient physical separation and strict time limits on
responding to the question.}.  The game being played (minority or anti-minority)---and thus the payoff matrix---is determined by the set of questions asked
by the referee. If the referee has asked three of the players for the value of
$Z$ and one of the players for the value of $X$, then the players are playing
the minority game, and the payoff matrix is the same as for the standard
minority game (independent of which player has been asked which particular question). If the referee asks three of the players for the value of $X$
and one player for the value of $Z$, then the payoff matrix is that of the
anti-minority game.  The referee has chosen the question list uniformly at random from the following chart before the game begins:
\begin{eqnarray*}
 \left. \begin{array}{c}
    X_1 Z_2 Z_3 Z_4\\
    Z_1 X_2 Z_3 Z_4\\
    Z_1 Z_2 X_3 Z_4\\
    Z_1 Z_2 Z_3 X_4
  \end{array} \right\} \hspace{0.75em} & \text{minority game;} \\
 \left. \begin{array}{c}
    Z_1 X_2 X_3 X_4\\
    X_1 Z_2 X_3 X_4\\
    X_1 X_2 Z_3 X_4\\
    X_1 X_2 X_3 Z_4
  \end{array} \right\} \hspace{0.75em} & \text{anti-minority game.}
\end{eqnarray*}
Thus, each of these lists has probability~$\tfrac 1 8$ of being asked by the referee.  Other question lists (such as $Z_1 Z_2 X_3 X_4$
or $X_1 X_2 X_3 X_4$) are promised not to be used.   Notice that once the payoff matrix is fixed (by the total number of each question
asked), there is no further dependence on which player
was asked which question, with the payoff determined entirely by the players'
answers (of $\pm 1$).

The list $X_1 Z_2 Z_3 Z_4$, for instance,
represents player~1 being asked for the value of $X$ and players 2--4 being
asked for the value of $Z$.  According to the chart, this corresponds to
the minority game.  Now let's say, for example, that player~3 answers~$+1$, while
players~1, 2, and~4 answer~$-1$.  Then player~3 receives a payoff of~1, and
the others receive nothing.  It does not matter which question player~3 was asked (in this case, ``What
is the value of~$Z$?''), only that his answer ($+1$) is different from the others' ($-1$) and that the game being played (as determined by all four questions together) is the
minority game.

\section{Analysis of the game}

In devising a strategy for this overall game, the challenge, of course, is
that the players don't know \tmtextit{a priori} whether they are playing the
(competitive) minority game or the (cooperative) anti-minority game. Since all
eight possibilities have equal probabilities, the two games are equally
likely.

The players gain partial information about the game being played, however,
once they receive their individual questions. If player~2, for example, is
asked for the value of~$X$, then he knows that only four possible question
lists remain: $Z_1 X_2 X_3 X_4$, $X_1 X_2 Z_3 X_4$, and $X_1 X_2 X_3 Z_4$,
which correspond to the anti-minority game, as well as $Z_1 X_2 Z_3 Z_4$,
which corresponds to the minority game. This means that if a player is asked
for the value of $X$, he will believe himself to be playing the anti-minority
game with 75\%~certainty and the minority game with 25\%~certainty. Similarly,
being asked for the value of~$Z$ will reverse these probabilities: 75\%~for
the minority game, 25\%~for the anti-minority game.

In principle, the players can use this information when devising their
strategies. Notice that in all cases, one player will ``get it wrong'' in that
she will only assign 25\% to the actual game being played. One might suspect
that, especially if the actual game being played is the minority game, which
is competitive, this asymmetry in information about the game could be used by
the other three players to ensure that the odd player out is never allowed to
win. We have not proven this; we only use it to illustrate the fact that since
a competitive game is never completely ruled out for any player, the strategic
analysis is nontrivial.

\section{Bounds on expected payoff regardless of strategy}\label{sec:bounds}

At this point, it is useful to examine the bounds on a player's expected
payoff \tmtextit{regardless of strategy}. Whether such a payoff is achievable
by any particular strategy is a separate---and important---question that will
be addressed shortly. But there are a few things that can be stated about the
game that must hold for any strategy:
\begin{enumerate}
  \item \tmtextbf{The maximum total expected payoff (sum for all players)
  is~1. }This bound is saturated if and only if the players can guarantee that
  they will always produce a win condition when asked any of the eight
  possible question lists.
  
  \item \tmtextbf{If there exists a Nash-equilibrium strategy that achieves a
  given total expected payoff, then there also exists a symmetric Nash
  equilibrium~\footnote{In game-theoretic terms the equilibria we shall discuss are \emph{correlated} Nash equilibria~\cite{Aumann1987}
since we allow the players to share information before the game.
However, we shall continue to use the term `Nash equilibrium' for brevity
and to be consistent with the previous literature in quantum games,
where the distinction between Nash equilibria and correlated equilibria
has not been made.} at that total expected payoff. }
  There is nothing to distinguish
  the players before the game begins---they all have the same information, and
  they all factor into the game in the same way. Thus, if the payoffs differ
  for different players using a given Nash-equilibrium strategy~$S$, then a
  correlated strategy that permutes the players' labels in a uniformly random
  fashion (using, for example, resources shared before the game)
  and then employs~$S$ is necessarily a symmetric Nash equilibrium
  with the same total expected payoff.
  
  \item \tmtextbf{The maximum symmetric Nash-equilibrium expected payoff for
  each player is~$\frac{1}{4}$, which is also Pareto optimal. }This follows
  directly from the two points above.

\end{enumerate}
Note that in point~(3) we are not claiming that this optimal payoff is
achievable. We show below, however, that a strategy using quantum resources
can be employed to achieve an expected payoff of $\frac{1}{4}$ for each
player, and we subsequently show this strategy to be a Nash equilibrium. Thus,
by point~(3), this quantum strategy truly is Pareto optimal and overall ``the
best the players can do'' for this game. Following this, we show that any
strategy limited to classical resources---i.e., those whose statistical
properties can be reproduced using a local hidden variable model---always
performs \tmtextit{strictly worse} than this, thus demonstrating that quantum
resources are required for achieving the optimal outcome.

\section{Stabilizer formalism}

We can use the stabilizer formalism to create a strategy using quantum
resources that achieves the Pareto-optimal payoff of $\frac{1}{4}$ for each
player. The strategy involves the players preparing a four-qubit entangled
quantum state, one qubit of which is taken by each player to be used during
the game. Players each make projective measurements on their system in accord
with the question asked ($X$ or~$Z$) and record $\pm 1$ in accord with the
result. The win conditions for the two possible games can be seen to be
mutually exclusive and based on the product of the players' answers: $- 1$ for
the minority game, $+ 1$ for the anti-minority game.

A stabilizer state is defined as the $+ 1$-eigenstate of a set of commuting
observables.  The set of all such observables is called the \emph{stabilizer} of the state~\cite{Nielsen2000}.  In order for the players to guarantee a
win in the cases where the minority game is being played, measurements made in
accord with the questions asked must always multiply to~$- 1$. Such a state is
stabilized by the following four operators: $- X_1 Z_2 Z_3 Z_4$, $- Z_1 X_2
Z_3 Z_4$, $- Z_1 Z_2 X_3 Z_4$, and $- Z_1 Z_2 Z_3 X_4$. Notice that these
operators commute. Therefore they form a valid set of generators of a
stabilizer group, which specifies a unique pure state
\begin{eqnarray}
\label{eq:psidef}
  \ketbra{\psi} &=& \left( \frac{I - X_1 Z_2 Z_3 Z_4}{2} \right) \left( \frac{I - Z_1 X_2 Z_3 Z_4}{2} \right) \\ \nonumber
 && \times \left( \frac{I - Z_1 Z_2 X_3 Z_4}{2} \right) \left( \frac{I - Z_1 Z_2 Z_3 X_4}{2} \right).
\end{eqnarray}
The state $\ket{\psi}$ guarantees that there is a winner in each of the four
minority games, no matter which of these four versions is played. This state
is locally equivalent to the four-qubit cat state.

The full stabilizer contains more than the four operators listed above. It
also contains all products of these four operators. Multiplying together all
choices of three of these four operators reveals the following four operators
to also be part of the stabilizer: $Z_1 X_2 X_3 X_4$, $X_1 Z_2 X_3 X_4$, $X_1
X_2 Z_3 X_4$, and $X_1 X_2 X_3 Z_4$. These correspond to ensuring that the
players' answers multiply to $+ 1$ for each of the four variants of the
anti-minority game. Thus, the state $\ket{\psi}$ also guarantees the players
win if the anti-minority game is played.

No matter which of the eight question lists is asked, making measurements in
accord with the questions on the state $\ket{\psi}$ guarantees that there is a
winner for each of them. Furthermore, the state is symmetric under permutation
of the players' labels, so each player's expected payoff is $\frac{1}{4}$ when
this quantum strategy is used.

\section{Proof of Nash Equilibrium}

We now prove that this quantum strategy is a Nash equilibrium.  To do this, we must show that no player can do better by unilaterally choosing to do something other than make a projective measurement in accord with his question ($X$~or~$Z$) on the jointly prepared state~$\ket \psi$ from Eq.~\eqref{eq:psidef}.  What are the alternatives?  Since he must answer with one of two possible responses ($\pm 1$), the most general measurement he can perform is a two-outcome positive-operator-valued measurement (POVM)~\cite{Nielsen2000}.  This is more general than a projective measurement, but in the two-outcome case it is equivalent to adding bias and noise to such a measurement and can be simulated with classical randomness~\footnote{To see this, consider that any two-outcome POVM must consist of two elements of the form~$\{ E, I - E\}$, with $E = E^\dag \ge 0$.  Thus, the two elements are diagonal in the same basis.  Transforming to that basis, we have~$E_+ = a \ket \uparrow \bra \uparrow + b \ket \downarrow \bra \downarrow$ and $E_- = (1-a) \ket \uparrow \bra \uparrow + (1-b) \ket \downarrow \bra \downarrow$, with $0 \le b \le a \le 1$.  (If we want $a \le b$, we can just exchange the labels on $E_\pm$.)  A true projective measurement would have $a=1$ and $b=0$ so that $E_+ = \ket \uparrow \bra \uparrow$ and $E_- = \ket \downarrow \bra \downarrow$.  The statistics of the POVM depend only on the diagonal elements of the density operator when written in the eigenbasis of $E_\pm$.  Thus, they are effectively classical and can be simulated by projectively measuring in this basis and postprocessing the result with flips of weighted coins.  Specifically, notice that the probability of getting the outcome corresponding to~$E_+$ is~$p(+1) = p({+1} | {\uparrow}) p({\uparrow}) + p({+1} | {\downarrow}) p({\downarrow})$.  The conditional probabilities can be interpreted as the heads-probability for a weighted coin whose weight is conditioned on the outcome of a projective measurement in the $\{\ket \uparrow, \ket \downarrow\}$~basis.  The POVM can therefore be simulated by the following procedure.  First, projectively measure the qubit in this basis.  Next, prepare a weighted coin whose weighting depends on the outcome of that measurement: if the result is~$\ket \uparrow$, then use~$p(\text{heads}) = a$; if it's~$\ket \downarrow$, then use~$p(\text{heads}) = b$.  Flip this coin.  If the result is heads, return~$+1$.  Otherwise, return~$-1$.  The statistics of this coin flip are the same as that of the POVM.}.  Thus, we will only consider projective measurements as the possible alternative strategies for the player (since adding bias and noise will only reduce his expected payoff).

Instead of starting directly with~$\ket \psi$, the players prepare for themselves the initial state
$\ket{\psi_{\text {in}}} = ( \ket{0000} - i \ket{1111}) /\sqrt{2}$.  We now define the unitary operation~\footnote{The overall
factor of $i$ in Eq.~\eqref{eq:NE} has no physical effect and is included for notational
convenience only.}
\begin{eqnarray}
  \label{eq:NE} \op{s} = \frac{i}{\sqrt{2}} ( \op{Y} + \op{Z}) =
  \frac{i}{\sqrt{2}} \left( \begin{array}{cc}
    1 & - i\\
    i & - 1
  \end{array} \right),
\end{eqnarray}
which satisfies~$\op s \otimes \op s \otimes \op s \otimes \op s \ket {\psi_{\text{in}}} = \ket \psi$.  Now we can demonstrate that the strategy of each player applying $\hat s$ to his qubit, measuring according to the question asked ($X$~or~$Z$), and reporting the corresponding result 
is a symmetric Nash equilibrium.

Suppose that the first three players choose to follow this strategy, while the fourth, $D$, elects to measure in a different basis, which could depend on which question she is asked.  Without loss of generality, this is equivalent to acting with a question-dependent unitary~$\op s_X$ or~$\op s_Z$ followed by a projective measurement of~$X$ or~$Z$, respectively.  Any such unitary operator may be parameterized in general (up to an overall phase) by
\begin{eqnarray}
  \op{M} (\theta, \alpha, \beta) = \left( \begin{array}{cc}
    e^{i \alpha} \cos \frac{\theta}{2} & ie^{i \beta} \sin \frac{\theta}{2}\\
    ie^{- i \beta} \sin \frac{\theta}{2} & e^{- i \alpha} \cos
    \frac{\theta}{2}
  \end{array} \right),
\end{eqnarray}
where $\theta \in [0, \pi]$, and $\alpha, \beta \in (- \pi, \pi]$.
Note that $\op s = \op{M} ( \frac{\pi}{2}, \frac{\pi}{2}, - \frac{\pi}{2})$.  We can calculate the expected payoff for~$D$ for each of the question lists that could be asked of the players.  Interestingly, the payoff does not depend on the game being played but rather only on the question that~$D$ is asked.  If she is asked for~$Z$, then
\begin{align}
  \label{eq:payD1} \av{\$_D}_Z = \frac{1}{8} - \frac{1}{8} \cos (\alpha - \beta)
  \sin \theta,
\end{align}
and if she is asked for~$X$, then
\begin{align}
  \label{eq:payD2} \av{\$_D}_X &= \frac{1}{8} - \frac{1}{16} \bigl[\cos 2 \alpha (1 + \cos \theta) +
  \cos 2 \beta (1 - \cos \theta) \bigr].
\end{align}
By inspection, Eqs.~(\ref{eq:payD1}--\ref{eq:payD2}) both achieve their maximum value of
$\av{\$_D} = \frac{1}{4}$ by the (non-unique) choice of $\theta = \alpha = -
\beta = \frac{\pi}{2}$, that is, by choosing~$\op s_X = \op s_Z = \op s$. Since $D$~cannot improve
her payoff by a unilateral change of strategy away from $\op s$ regardless of which of
the games is being played or which question she is asked, this is her Nash-equilibrium strategy. By symmetry, the other
players also maximize their payoffs with the same strategy.  Indeed, by point~(3) in the Section~\ref{sec:bounds}, an average payoff of $\frac{1}{4}$ for each player is the maximum
possible and therefore Pareto optimal.

\section{Local Hidden Variable Model}

We now prove that no classical strategy (by which we mean a strategy admitting a local hidden
variable model) can reproduce the quantum mechanical payoff.  There are many possible classical strategies that may be employed. These
include both mixed strategies, which have an element of randomness, and pure
strategies, which are deterministic. However, mixed strategies can be replaced
by corresponding pure strategies with all randomness relegated to the
classical resource prepared before the game.

The players meet before the game to prepare a classical resource to take with
them to the game. This resource can be prepared stochastically, but the
existence of a local hidden-variable model (by assumption) guarantees that all
statistical properties of the resource can be modeled as arising from a joint
probability distribution~$p (\lambda_1, \lambda_2, \lambda_3, \lambda_4)$ over
a set of definite, classical resources~$\{\lambda_i \}$. This means that at
the end of the joint meeting, each player will always come away with a
definite classical resource~$\lambda_i$, even if the particular resource was
chosen partially using coin flips and spins of roulette wheels. This already
introduces a source of stochasticity into the problem, with deterministic
preparation being a special case.

Once at the game, each player is asked a question~$q_i$, after which each
player must make a choice to either report the answer~$+ 1$ or to report~$-
1$. Each may use a mixed strategy to do this. That is, each player can choose
also to bring with her coins for flipping and roulette wheels for spinning
\tmtextit{after} the question has been asked, in order to help her make a
decision of which value to report, but this is redundant. The players know the
possible questions before the game starts, and the existence of hidden
variable models means that all probabilities may be interpreted as lack of
knowledge of the real state of affairs of a system. Thus, each player can flip
a coin or spin a roulette wheel in the pre-game meeting and record the
outcomes for use in the game instead of generating these outcomes after the
fact. Thus, all strategies for player~$i$ that involve stochastic dependence
on the classical resource~$\lambda_i$ (i.e., those which involve additional
stochastic variables~$\Omega_i$) can be simulated using a strategy that
depends deterministically on a new classical resource~$\lambda'_i =
(\lambda_i, \Omega_i)$. (Note that we drop the prime in what follows.) Notice
that the existence of a local hidden-variable model for all statistical
aspects of the classical game is required for this replacement to be made---it
cannot in general be done when quantum resources (specifically, certain
entangled states) are used.

We can therefore consider each player's strategy to be defined by a
function~$f_i$, which defines a pure strategy such that the reply to the
question $q_i$ is determined by $f_i (\lambda, q_i)$, which---in a slight
abuse of notation---we shorten to $q_i (\lambda_i)$ for clarity.  By point~(1) in Section~\ref{sec:bounds}, simulating the Pareto-optimal
Nash-equilibrium quantum strategy where each player has an expected payoff of
$\frac{1}{4}$ requires that the players be able to guarantee a winner in any
of the eight cases that could be posed to them. This condition can be written
as a set of simultaneous equations:
\begin{subequations}
\label{eq:LHVeqs}
\begin{align}
X_1(\lambda_1) Z_2(\lambda_2) Z_3(\lambda_3) Z_4(\lambda_4) &= -1,\\
Z_1(\lambda_1) X_2(\lambda_2) Z_3(\lambda_3) Z_4(\lambda_4) &= -1,\\
Z_1(\lambda_1) Z_2(\lambda_2) X_3(\lambda_3) Z_4(\lambda_4) &= -1,\\
Z_1(\lambda_1) Z_2(\lambda_2) Z_3(\lambda_3) X_4(\lambda_4) &= -1,\\
Z_1(\lambda_1) X_2(\lambda_2) X_3(\lambda_3) X_4(\lambda_4) &= +1,\\
X_1(\lambda_1) Z_2(\lambda_2) X_3(\lambda_3) X_4(\lambda_4) &= +1,\\
X_1(\lambda_1) X_2(\lambda_2) Z_3(\lambda_3) X_4(\lambda_4) &= +1,\\
X_1(\lambda_1) X_2(\lambda_2) X_3(\lambda_3) Z_4(\lambda_4) &= +1.
\end{align}
\end{subequations}
Since~$q_i (\lambda_i)$ is a
deterministic function, it must take on a single value regardless of which
equation it appears in. Straightforward calculation reveals that there is no
solution to these equations. For instance, since each function must
return~$\pm 1$, multiplying equations~(a), (b), and~(c) gives
\begin{eqnarray*}
  X_1 (\lambda_1) X_2 (\lambda_2) X_3 (\lambda_3) Z_4 (\lambda_4) = - 1,
\end{eqnarray*}
in direct contradiction to~(h). Hence, no classical resource (either
deterministic or stochastic, using the argument above) can be prepared that
guarantees there is always a winner of each possible game. This means that,
using only classical resources, $\av{\$} < \frac{1}{4}$, where the inequality
is strict. Thus, there is no classical strategy which reproduces the result
that can be achieved using quantum resources.

\section{Bounds on expected payoff using only classical
resources}\label{sec:classbounds}

Having proved the main point of the paper, we could stop there, but we will
instead go one step further by demonstrating a finite gap between the maximum
symmetric expected payoff of~$\frac{1}{4}$ in the quantum case and that in the
classical case. This bound will be derived independently of any particular
classical strategy used. It is thus not known whether it is achievable or
whether it is a Nash equilibrium. The only purpose is to quantify how good one
could ever hope to do with only classical resources.

Consider first the fact that once a player has been asked his question, the
list of eight equations to be satisfied shrinks to four. These always have the
form of three instances of one game and one instance of the other. For
instance, if player~4 is asked for the value of~$Z$, then Eqs.~(\ref{eq:LHVeqs}d--g) are
eliminated. Unfortunately, as shown in the last section, the remaining four
cannot be satisfied simultaneously. The two sets are exchanged if she is asked
for the value of~$X$, but this doesn't help because Eqs.~(\ref{eq:LHVeqs}d--g) have no
solution either. The game is symmetric under exchange of the players' labels,
so this applies to all players equally. Thus, once the players know their
questions, each of them is left with a (different) set of four simultaneously
insoluble equations that cannot be ruled out.

Thus, it is not rational (even in the case of players with very clever
strategies!) for any player to believe that there will always be a winner in
each of the four cases he must still consider. Thus, he is forced to conclude
that there will be no winner in at least one of those games, even in the best
possible scenario. The question he must ask then is, \tmtextit{Which win am I
willing to sacrifice?} The answer, of course, depends on how well he expects
to do in each case.

Again, let's push the envelope as far as we can, while still remaining
rational. There are two win conditions that must be considered: a win of the
anti-minority game always pays~$\frac{1}{4}$, while a win of the minority game
has an expected payoff anywhere between~0 and~1, since there remains a
question of which player is the winner. The minority game win is thus the case
that requires more consideration. If a player considers the case where a
minority game is being played, she knows that three players have the same
information (they have each been asked for the value of~$Z$), while one player
has other information (he has been asked for the value of~$X$). Being
rational, the players who were asked~$Z$ must have equal expected
payoffs~$M_Z$, since there is nothing to distinguish between them, but it need
not be the same as the expected payoff~$M_X$ of the player who was asked
for~$X$. Still, the total payoff of the minority game is~1, so $3 M_Z + M_X \le 1$
when the minority game is being considered. We are considering the best
possible case for the players, so we choose to saturate this bound and write
\begin{equation}
\label{eq:minpayoff}
	3M_Z+M_X = 1,
\end{equation}
since any other choice only reduces the players' payoffs. Notice that $0 \le
M_Z \le \frac{1}{3}$, while $0 \le M_X \le 1$. The upper bound of~1 on~$M_X$
corresponds to the notion that a very clever player might have a strategy that
lets him win every time the minority game is being played and he is asked
for~$X$, while $M_Z$ is upper-bounded by~$\frac{1}{3}$ because there are three
players who are asked~$Z$ in the case of a minority game, so the full payoff
must have an equal chance of being obtained by any of them.

Let us consider now the case of a player who has been asked for the value
of~$X$. There is now one possible way that a minority game is being played and
three possible ways that an anti-minority game is being played. She knows that
she cannot guarantee a win in all four cases, yet each is equally likely. She
now has to choose between allowing a loss (for all players) in the minority
game and allowing a loss in one of the anti-minority games. Sacrificing the
minority game gives an expected payoff of
\begin{equation}
{\av{\$}}_{X,m} = \frac{3}{4} \, {\frac{1}{4}} + {\frac{1}{4}} \, 0 = {\frac{3}{16}},
\end{equation}
while sacrificing one of the three anti-minority games gives
\begin{equation}
{\av{\$}}_{X,a} = {\frac{3}{4}} \, {\frac{2}{3}} \, {\frac{1}{4}} + {\frac{1}{4}} \, M_X = {\frac{1+2M_X}{8}}.
\end{equation}

For this case, which game to sacrifice depends on the expected payoff~$M_X$ in
the case of a win condition in the minority game. When $M_X > \frac{1}{4}$,
then $\av{\$}_{X, a} > \av{\$}_{X, m}$, and the player is better off using a
strategy that sacrifices the win in the minority game. However, when $M_X <
\frac{1}{4}$, the player is better off sacrificing one of the anti-minority
game wins.

Similarly, if a player is asked for the value of~$Z$, there are three possible
ways that the minority game is being played and only one possible way that the
anti-minority game is being played, each again being equally likely. He must
choose which one to allow a loss in. Sacrificing one of the minority games
gives
\begin{equation}
{\av{\$}}_{Z,m} = {\frac{1}{4}} \, {\frac{1}{4}} + {\frac{3}{4}} \, {\frac{2}{3}} M_Z = {\frac{1+8M_Z}{16}} = {\frac{11-8M_X}{48}},
\end{equation}
where we have used \eqref{eq:minpayoff} in the last equality. Instead, sacrificing the
anti-minority game gives
\begin{equation}
{\av{\$}}_{Z,a} = {\frac{1}{4}} \, 0 + {\frac{3}{4}} \, M_Z = {\frac{3M_Z}{4}} = {\frac{1-M_X}{4}}.
\end{equation}
Once again, the cross-over point between the two occurs when $M_X = \frac{1}{4}$,
with a sacrifice of the minority game preferred when $M_X < \frac{1}{4}$
and sacrificing the anti-minority game preferred when $M_X > \frac{1}{4}$.

The overall expected payoff is an average of the two optimal payoffs. We
consider the two cases of~$M_X$ separately. When $M_X < \frac{1}{4}$, then
\begin{equation}
\label{eq:maxsmallMX}
\max {\av{\$}} = {\frac{1}{2}}{\av{\$}}_{X,m} + {\frac{1}{2}}{\av{\$}}_{Z,m} = {\frac{7-4M_X}{32}}.
\end{equation}
Similarly, when $M_X > \frac{1}{4}$, then
\begin{equation}
\label{eq:maxbigMX}
\max {\av{\$}} = {\frac{1}{2}}{\av{\$}}_{X,a} + {\frac{1}{2}}{\av{\$}}_{Z,a} = {\frac{17+4M_X}{96}}.
\end{equation}
Recalling the bounds on~$M_X$, plugging~$M_X = 0$ into \eqref{eq:maxsmallMX} 
and plugging~$M_X = 1$ into \eqref{eq:maxbigMX} both give
\begin{equation}
\max {\av{\$}}_{\text{classical}}={\frac{7}{32}}.
\end{equation}
Once again, we have not shown that this is achievable by any particular
strategy nor that such a strategy is a Nash equilibrium. However, if a symmetric
Nash-equilibrium strategy is to exist for this game under the restriction that
the players only use classical resources, the best any player can hope for is
an expected payoff of~$\frac{7}{32}$, instead of~$\frac{1}{4}$ using quantum
resources.

\section{Conclusion}

By selecting either a minority or anti-minority game and choosing one of four
measurement bases in each case, we have constructed a competitive game with an
average Nash-equilibrium payoff that is also Pareto optimal. There is no
classical prescription---that is, a set of values for measurements of~$X_i$
and~$Z_i$---that can produce a winner in all four minority and all four
anti-minority games. Thus, no classical strategy, even allowing for initial
classical communication amongst the players, can achieve the payoff that we
have demonstrated using quantum entanglement. The best conceivable average
payoff using only classical resources is only $\frac{7}{8}$ of that achievable
by distributing an entangled quantum state amongst the players.

Although this game is somewhat contrived, it answers the two main criticisms
of quantum games presented in Ref.~{\onlinecite{vanEnk2002}}: (1) that the ``quantum
solution'' to a classical game does not faithfully solve the original
classical game, and (2) that the quantum solution can also be obtained through
classical means. Our responses follow. We also address the ``side issue''
presented in that paper involving the cooperative aspect of games, which we
discuss separately below.

In response to the first objection, the quantum and classical versions of the
game we consider are \tmtextit{identical} except for the resources that are
available to the players in each case. In the end, each player must produce a
definite classical answer ($\pm 1$) for the referee, even if quantum resources
were used to make the decision---i.e., the players answer with
\tmtextit{bits}, not qubits. This looks to be as close as one could possibly
get to a ``minimal generalization'' of a classical game to a quantum one:
simply replace classical resources with quantum ones while requiring the same
rules and interactions with the referee. This generalization from classical to
quantum is more conservative than the ``standard'' method of quantizing a
classical game presented in Ref.~{\onlinecite{Eisert1999}} since the quantum
resources (if any) possessed by each player can only be manipulated
\tmtextit{locally} before a classical answer is produced. As such, many of the
objections to the method of Ref.~{\onlinecite{Eisert1999}} do not apply to our game.
The second objection is addressed directly in Section~\ref{sec:classbounds},
where we prove that players of our game perform \tmtextit{strictly better}
when they are able to use quantum resources than when they are restricted to
classical ones.

It is worth noting that the only difference
between the games we consider is whether the players are capable of using
quantum resources in addition to classical ones or not. All other aspects of
the game---such as the availability of correlated randomness, the ability to
make (and~break!) agreements, etc.---remain unchanged. In fact, the inability
to use quantum resources can be framed in physical terms as, for example,
ignorance of the fact that the world is quantum mechanical (e.g., if the game
were played in the year 1900) or insufficient technical expertise in preparing
and manipulating quantum resources. This means that we really just have
\tmtextit{one game} that we are analyzing, with the two ``versions'' of it
arising simply from two different scenarios regarding the knowledge and
abilities of the players.

The final issue in Ref.~{\onlinecite{vanEnk2002}} is whether the quantum version of
a classical game has made an inherently non-cooperative game into a game with a
cooperative aspect.  Our response is that the degree to which our game is,
respectively, ``cooperative'' and ``competitive'' does not change when
comparing the classical and quantum versions since the game itself has not
changed.  It is worth emphasizing a related point, however: the issue of competition.  The implied skepticism in Ref.~\onlinecite{vanEnk2002} about whether entanglement can help the players improve their payoff in a wholly noncooperative game is well taken, and we note that it is the uncertainty in whether it is best to compete or cooperate that appears to make the improvement possible in the case we consider.  While our game retains a cooperative aspect, it also involves competition between the players.  This is the key feature that distinguishes it from others in the literature that are wholly cooperative~\cite{Cleve2004,Brassard2005}: entangled quantum resources assist the players in improving their expected payoff \emph{in a competitive setting}.  Thus, one might conjecture that uncertainty about cooperation\slash competition is a generic requirement for allowing entanglement to assist with a competitive game (beyond what is achievable using classical correlations).  We leave this question to further research.

\begin{acknowledgments}
The authors are grateful for useful comments by
David Pritchard, University of Waterloo
and Steven van~Enk, University of Oregon.
This work was supported by the Australian Research Council, the Australian
Government, and the US National Security Agency~(NSA) and the Army Research
Office~(ARO) under contract number W911NF-08-1-0527. Research at Perimeter
Institute is supported by the Government of Canada through Industry Canada and
by the Province of Ontario through the Ministry of Research \& Innovation.
\end{acknowledgments}

\bibliographystyle{bibstyleNCM}
\bibliography{hill2009v2}

\end{document}